\documentclass[epj,preprint]{svjour}
\usepackage{graphicx}
\usepackage{bm}
\usepackage{epsfig}
\usepackage{amssymb}
\usepackage{amsmath}
\usepackage{bbm}
\usepackage[normalem]{ulem}
\newcommand{\rd}{\ensuremath{\hbox{d}}}
\newcommand{\id}{\ensuremath{\,\rd}}
\newcommand{\bra}[1]{\langle#1\rvert}
\newcommand{\ket}[1]{\lvert#1\rangle}

\newcommand{\bracket}[3]{\bra{#1}#2\ket{#3}}

\newcommand{\be}{\begin{equation}}
\newcommand{\ee}{\end{equation}}
\newcommand{\ba}{\begin{eqnarray}}
\newcommand{\ea}{\end{eqnarray}}
\newcommand{\bi}{\begin{itemize}}
\newcommand{\ei}{\end{itemize}}

\newcommand{\dd}[2]{\ensuremath{\frac{\mathrm{d}#1}{\mathrm{d}#2}}}

\begin{document}
\title{The Similarity Renormalization Group for
  Three-Body Interactions in One Dimension}

\author{O. \r{A}kerlund, E.~J. Lindgren,  J. Bergsten, B. Grevholm,  P. Lerner, R.
  Linscott, C. Forss\'en, L. Platter}
\authorrunning{O. \r{A}kerlund {\it et al.}}
\institute{ Chalmers University of Technology, Department of Fundamental Physics,
              SE--412 96 Gothenburg, Sweden}
\date{\today}
%

\abstract{ We report on recent progress of the implementation of the
  similarity renormalization group (SRG) for three-body interactions
  in a one-dimensional, bosonic model system using the plane wave
  basis. We discuss our implementation of the flow equations and show
  results that confirm that results in the three-body sector remain
  unchanged by the transformation of the Hamiltonian. We also show how
  the SRG transformation decouples low- from high-momentum nodes in
  the three-body sector and therefore simplifies the numerical
  calculation of observables.
} 
\maketitle
%
\section{Introduction\label{sec:intro}}
Renormalization group methods have become an important tool in modern
physics \cite{rg-royal}. In particular, for studies of strongly-interacting
many-body systems they frequently facilitate the correct calculation
of observables. One of these methods is the similarity renormalization
group (SRG), which has been extensively used in condensed matter
physics \cite{book:kehrein} but has recently become of importance also
in nuclear physics \cite{Bogner:2006pc,Bogner:2009bt}. The SRG
essentially constitutes a flow equation generating unitarily
equivalent Hamiltonians, which after evolution possess certain
features that usually make the calculation of observables 
easier. The form of the evolved potential will depend on the so-called
generator that is an essential ingredient of the flow
equation. Frequently used generators drive the two-body potential to
the diagonal in momentum space. Off-diagonal elements are thus driven
to zero and low momenta effectively decouple from large momenta.

Such capabilities are very important in ab initio nuclear structure
physics for which high-performance computing has become of increasing
importance and the available computational resources set a hard limit
on the number of observables that can be
calculated~\cite{Vary:2009qp}. Recently, the SRG flow equations for
three-body interactions were implemented for the truncated harmonic
oscillator basis and first nuclear structure calculations with
consistently evolved two- and three-body interactions have emerged
\cite{Jurgenson:2009qs,Jurgenson:2010wy,Roth:2011ar}. The SRG
therefore seems to provide a way to extend the limits of possible
computations. In these calculations the interactions were evolved in
the harmonic oscillator basis. However, a calculation of the evolved
three-body interaction in the plane wave basis would not only
facilitate the projection of the potential on any basis but also the
calculation of {\it infinite} matter observables such as the energy
per particle of nuclear matter. In
Refs.~\cite{Bogner:2005sn,Hebeler:2010xb} it was shown that an evolved
two-body interaction in combination with a phenomenological three-body
interaction seems to lead to a converging many-body perturbation
theory series for infinite matter. However, a missing link in this
calculation remains the inclusion of a consistently evolved three-body
interaction. Instead, parameters of the leading chiral effective field
theory three-nucleon force were refitted and used alongside the
evolved two-body potential. Since these positive results for light
nuclei and nuclear matter validate also an effort to construct an ab
initio density functional based on many-body perturbation theory with
evolved interactions \cite{Drut:2009ce}, the SRG seems therefore to
have impact on calculations across the whole chart of nuclides.

In this work we consider the evolution of a one-dimen\-sional
Hamiltonian with two- and three-body interaction terms.  In
Sec.~\ref{sec:simil-renorm-group} we will introduce the flow equations
that determine the evolution of the Hamiltonian. In the following two
sections we will then discuss how the flow equations are implemented
for two- and three-body interactions in bosonic systems. The
presentation of numerical results will focus on illustrating key
properties of evolved interactions, i.e. the conservation of observables
and (for our choice of generator) the effective decoupling of low and high momenta.
In the last section we summarize our results and discuss necessary steps
towards an extension of this work to three dimensions and to nuclear
systems. 
\section{The Similarity Renormalization Group}
\label{sec:simil-renorm-group}
Let us denote with $H_s$ the transformed (and initially unknown)
Hamiltonian where $s$ denotes the so-called flow parameter and
provides a measure of how much the Hamiltonian has been
transformed. Then, there exists a unitary transformation $U_s$ such that
\begin{equation}
H_s=U_sHU_s^{\dagger},
\end{equation}
where $H=H_{s=0}$ is the known, original Hamiltonian. We can now calculate
the derivative of the above equation with respect to $s$
\begin{equation}
\dd{H_s}{s} =\dd{U_s}{s}HU_s^{\dagger}+U_sH\dd{U_s^{\dagger}}{s}\label{f1}~.
\end{equation}
Since the transformation $U_s$ is unitary we have $U_sU_s^{\dagger}=1$
and therefore
\begin{eqnarray}
\dd{U_s}{s}U_s^{\dagger}=-U_s\dd{U_s^{\dagger}}{s}\equiv\eta_s\label{f2}~.
\end{eqnarray}
Using this in Eq.~\eqref{f1} gives
\begin{eqnarray}
\dd{H_s}{s}&=&\eta_sH_s-H_s\eta_s = [\eta_s,H_s] \label{eq:srg_def}.
\end{eqnarray}
We can then specify the unitary transformation by specifying $\eta_s$,
which is subject to the condition
\begin{equation}
\label{ec}
\eta_s^{\dagger}=-\eta_s~,
\end{equation}
which follows from Eq.~\eqref{f2}.

A convenient choice of $\eta_s$ is $\eta_s=[G_s,H_s]$ where $G_s$ is a
Hermitian operator. It obeys Eq.~\eqref{ec} since $H_s$ is also
Hermitian:
\begin{equation}
\left[G_s,H_s\right]^{\dagger}=-[G_s,H_s]~.\\
\end{equation}
We will store all dependence on the flow parameter  $s$ in the potential term of the
Hamiltonian, thus writing $H_s=T_{\text{rel}}+V_s$, where $T_{\rm{rel}}$ is
the relative kinetic energy operator. Expanding the commutators then
gives the equation
\begin{equation}
\label{eq:srg-general}
\dd{H_s}{s} = \dd{V_s}{s}=G_sH_sH_s+H_sH_sG_s-2H_sG_sH_s.
\end{equation}
There is significant freedom in the choice of the generator $G_s$.  In
this work we have chosen $G_s=T_{\rm{rel}}$.  This generator is used
since it is known to drive the Hamiltonian to the diagonal, which is
usually desirable. The reason for this feature is that $T_{\rm{rel}}$
is, in itself, diagonal in momentum space.

\section{Momentum Space Equations}
\label{sec:interactions}
We will define the interactions with respect to the usual three-body
Jacobi momenta
\begin{eqnarray}
  \label{eq:jacobi}
\nonumber
  {\bf p}&=&\frac{1}{2}\left({\bf k}_1-{\bf k}_2\right)~,\\
  {\bf q}&=&\frac{2}{3}\left( {\bf k}_3-\frac{1}{2}({\bf k}_1+{\bf k}_2)\right)~.
\end{eqnarray}
We will be working in a partial-wave projected basis. This needs to be
clarified since in one dimension plane waves can only propagate in two
directions; backwards and forward. In one dimension, there exist
therefore only two partial waves $l=0$ and $l=1$, which correspond also
to the parity of the state. We will be interested in bosons in this
work and therefore assume that the two-body system is symmetric under
exchange of the particles. Thus, we will only work with $l=0$ basis states.
In the two-body sector, a complete set of states will therefore be
written as
\begin{eqnarray}
  \label{eq:2cst}
  1=\int_0^\infty \hbox{d}p\;|p \:l=0\rangle\langle p \: l=0|=\int \hbox{d}p\;|p\:
  0\rangle\langle p\: 0|~.
\end{eqnarray}
In the three-body sector we will concentrate on states with total
angular momentum $L=0$, which implies that the relative angular
momentum associated with the $q$ variable is $\lambda=0$. The complete
set of states in the three-body sector is therefore
\begin{eqnarray}
  \label{eq:3cst}
  1=\int_0^\infty \hbox{d}p\int_0^\infty\hbox{d}q\;|p\,q \:(00)0\rangle\langle
  p\,q \:(00)0|~,
\end{eqnarray}
where the parentheses denote that two angular momenta have been coupled
to total angular momentum $L$. From now on we will drop all angular
momentum information in the bras and kets since we have only one angular momentum
state in the two- and three-body system, respectively.

Another quantity useful to define is the hypermomentum $\zeta$,
defined by
\begin{eqnarray}
  \label{eq:hypermom}
  \zeta^2=p^2+\frac{3}{4}q^2~.
\end{eqnarray}
Since this quantity is proportional to the total, relative
kinetic energy of the three-body system it also defines a plane in
which the three-body interaction will become diagonal through
evolution.

\subsection{Interactions}
\paragraph{Two-Body Potentials:}
\label{sec:two-body-potential}
We have used different two-body potentials in this work to analyze the
features of the SRG evolution. In particular, we have used a separable
potential that has the advantage that the binding energy and the
two-body t-matrix can be calculated analytically
\begin{eqnarray}
  \label{eq:separable}
  V_\mathrm{sep}(p,p')&=&g \exp(-p^2/\Lambda^2)\exp(-p'^2/\Lambda^2)~.
\end{eqnarray}
As the regulator $\Lambda$ is increased towards infinity the potential takes
the form of a delta-function in coordinate space. In this case the
binding energy of the $N$-boson state is known analytically and
provides an excellent test for our few-body code. For additional benchmarking of
numerical results we have also employed a potential that was
previously used in Ref.~\cite{Jurgenson:2008jp}
\begin{equation}
V(p,p') = \sum_{i=1,2}
\frac{V_i}{2\pi}\exp{\left(-\frac{(p-p')^2\sigma_i^2}{4}\right)},
\label{eq:Valpha}
\end{equation}
with parameters given in Table~\ref{tab:jurgparam}. It is important to
note that we employ the {\it partial-wave} projected versions of the
above potentials.
\begin{table}[t]
\begin{center}
\caption{Parameters for the two-body potential given by
  Eq.~\eqref{eq:Valpha}.}
\label{tab:jurgparam}
\begin{tabular}{c|cccc}
&$V_1$ & $V_2$ & $\sigma_1$ & $\sigma_2$ \\ \hline
$V_{\alpha}$&12 &  -12 & 0.2 & 0.8\\
$V_{\beta}$&0 &  -2 & 0 & 0.8\\
\end{tabular}
\end{center}
\end{table}
The parameters in Table \ref{tab:jurgparam} show that we will use a
purely attractive interaction ($V_\beta$) and an attractive interaction with {\it
  short-range} repulsion ($V_\alpha$).
\paragraph{The Three-Body Potential:}
The SRG evolution will generally induce many-body forces, however, it
is expected that the three-body force will dominate over higher
many-body forces as long as the flow parameter $s$ is not too
large. To complete our analysis, we also added a three-body force to
the unevolved Hamiltonian to mimic general features present in nuclear
physics. The three-body potential that we have used is of the same
simple form as the one used in Ref. ~\cite{Jurgenson:2008jp}
\begin{equation}
  \label{eq:3body}
  V_3(p,q,p',q') = \sqrt{3} c_E f_{\Lambda}(p,q) f_{\Lambda}(p',q'),
\end{equation}
where $c_E$ is the strength of the interaction and 
\begin{equation}
f_{\Lambda}(p,q) = \exp{\left(-\left(\frac{(2p^2+\tfrac{3}{2}q^2)}{\Lambda^2}\right)^n\right)}~,
\end{equation}
where we use $n=4$ and $\Lambda=2$ throughout this work.
\subsection{SRG Equations}
\paragraph{Two-Body SRG:}
With the kinetic energy operator as generator for the SRG equation,
Eq.~\eqref{eq:srg-general} becomes
\begin{equation}
\frac{\rd V_s}{\rd s} = 2TV_sT + V_sV_sT + TV_sV_s - V_sTT - TTV_s -
2V_STV_s .
\end{equation}
Using the complete set of states defined in Eq.~\eqref{eq:2cst} we can
write out the evolution equations in momentum space
\begin{eqnarray}
\label{eq:two_body_momspace}
\nonumber
\frac{\rd}{\rd s}\bracket{p}{V_s}{p'} &=& -\left( p^2-p'^2 \right )^2
\bracket{p}{V_s}{p'}\\ 
\nonumber
&&\hspace{-1cm}+ \left ( p^2+p'^2 \right )\int \id q
\bracket{p}{V_s}{q}\bracket{q}{V_s}{p'}\\
&&\hspace{-0cm} -2\int \id q q^2 \bracket{p}{V_s}{q}\bracket{q}{V_s}{p'}~.
\end{eqnarray}
The first term on the right hand side ensures that the potential is
driven to the diagonal.
\paragraph{Three-Body SRG:}
The equations for the evolution of the three-body potential are
significantly more complicated. We write the Hamiltonian in the three-body
sector as
\begin{equation}
H_s = T + V_2^{(1)} + V_2^{(2)} + V_2^{(3)} + V_3 ~,
\label{eq:Ham1}
\end{equation}
where $V_2^{(i)}$ denotes the two-body potential in the three
different channels. We have also dropped the subscript $s$ from the
potential and will keep doing this from now on to simplify
notation. Since we are working with identical bosons we will assume
that the induced three-body interaction is symmetric under the
exchange of two particles \footnote{In the fermionic case we will
  write $V_3=V_3^{(1)} + V_3^{(2)} + V_3^{(3)}$}. The flow equation in
the three-body sector is then written as
\begin{equation}
\frac{\rd}{\rd s}(V_2^{(1)} + V_2^{(2)} + V_2^{(3)} + V_3) = 
\left[\left[T,H_s\right],H_s\right].
\label{eq:flow-2and3body}
\end{equation}
The expression above contains the combined evolution of the two- and
three-body interaction. It is of general interest to separate the
two-body from the three-body evolution but in this case it removes also
spectator $\delta$-functions that arise from disconnected diagrams and
complicate the numerical computation of the evolved
potential. Following \cite{Bogner:2006pc}, we
circumvent this issue by subtracting from the above expression the
evolution of the two-body potentials $V_2^{(i)}$. This isolates the
derivative of the three-body potential and removes afore\-mentioned
$\delta$-functions
\begin{equation}
\frac{\rd V_3}{\rd s} = \left[\left[T,H_s\right],H_s\right] - 
\sum\limits_{i=1}^3 \frac{\rd V_2^{(i)}}{\rd s}.
\label{eq:flow-3body}
\end{equation}
Expanding the commutators
and rewriting the two-body differential equations as in previous
section gives us
\begin{equation}
\label{eq:srg3body}
 \frac{\rd V_3}{\rd s}   =  \mathcal{O}_2+\mathcal{O}_{23}+\mathcal{O}_3~,
\end{equation}
where we have defined
\begin{eqnarray}
\nonumber
\mathcal{O}_2 &=& \displaystyle{\sum_{i,j=1}^3}(1-\delta_{ij})\left(TV_2^{(i)}V_2^{(j)} + V_2^{(i)}V_2^{(j)}T - 2V_2^{(i)}TV_2^{(j)}\right),\\
\nonumber
\mathcal{O}_{23} &=& \displaystyle{\sum_{i=1}^3}\bigg(TV_3V_2^{(i)} + V_3V_2^{(i)}T - 2V_3TV_2^{(i)} +\nonumber\\
\nonumber
&& \quad\quad\qquad+ TV_2^{(i)}V_3 + V_2^{(i)}V_3T - 2V_2^{(i)}TV_3\bigg),\\
\nonumber
\mathcal{O}_3 &=& 2TV_3T - 2V_3TV_3 + \\
&&\hspace{2cm}TV_3V_3 + V_3V_3T - V_3TT - TTV_3.
\end{eqnarray}
We can express the two-body potentials $V_2^{(2)}$ and $V_2^{(3)}$
through the potential $V_2^{(1)}$ after application of permutation
operators 
\begin{eqnarray}
\nonumber
V_2^{(2)} &=& P_{13}P_{23}V_2^{(1)}P_{12}P_{23}~,\\
V_2^{(3)} &=& P_{12}P_{23}V_2^{(1)}P_{13}P_{23}~.
\end{eqnarray}
At this point it is useful to define the operator $P$ given by
\begin{eqnarray}
  \label{eq:perm}
  P=P_{12}P_{23}+P_{13}P_{23}~,
\end{eqnarray}
where $P_{ij}$ denotes the permutation operator that exchanges
particles $i$ and $j$. It can be shown that the overlap matrix
elements for $P_{12}P_{23}$ and $P_{13}P_{23}$ in a partial-wave
projected basis are identical. We can therefore write
\begin{eqnarray}
\nonumber
V_2^{(2)} &=& \frac{1}{4}P V_2^{(1)}P~,\\
V_2^{(3)} &=& \frac{1}{4}P V_2^{(1)}P~,
\end{eqnarray}
which simplifies the above equations significantly. 

We have used two different representations of the matrix element of
the operator $P$
\begin{eqnarray}
\nonumber
\langle pq|P|p'q'\rangle &=&
\displaystyle{\sum_{x=\pm1}}\delta(p-\pi(q,q',x))\delta(p'-\pi(q',q,x)),\\
\nonumber
\langle pq|P|p'q'\rangle &=&\displaystyle{\sum_{x=\pm1}}\delta(p-\tilde{\pi}(p',q',x))\delta(q-\chi(p',q',x)),\\
\label{eq:permute}
\end{eqnarray}
where 
\begin{eqnarray}
\nonumber
\pi(q,q',x)&=&\sqrt{\frac{1}{4}q^2+q'^2+xqq'},\\
\nonumber
\tilde{\pi}(p,q,x)&=&\sqrt{\frac{1}{4}p^2+\frac{9}{16}q^2+\frac{3}{4}p
q x},\\
\chi(p,q,x)&=&\sqrt{p^2+\frac{1}{4}q^2 - p q x}.
\end{eqnarray}

The obvious consequence of the implementation of the operator $P$ is
off-grid momenta in the object it is applied on. This problem can be
solved by splining these objects, e.g. a function containing the
shifted momentum $\pi(q,q',x)$ will be written as
\begin{eqnarray}
  \label{eq:spline}
  f(\pi(q,q',x))=\sum_i^N S_i(\pi(q,q',x))f(q_i)~.
\end{eqnarray}
We have used the global splines defined in Ref.~\cite{splines} but
also the cubic splines given in Ref.~\cite{Huber:1996td}. While the cubic splines provide
a speedup in the calculation they also decrease the accuracy of
results slightly. The results in this work were therefore all obtained
with global splines.

\section{Observables}
\label{sec:observables} 
\subsection{Two-Body Observables}
\label{sec:two-body-observables}
\paragraph{Phaseshifts and binding energies:}
We have calculated scattering and bound-state properties in the
two-body sector. Scattering properties are obtained by solving the
Lippmann-Schwinger equation. In operator form it is given by
\begin{eqnarray}
  \label{eq:LSeq}
  t=V+VG_0(E) t~,
\end{eqnarray}
where $G_0(E)$ denotes the free Green's function
\begin{eqnarray}
  \label{eq:G0}
  G_0(E) = \frac{1}{E-p^2/m+i\varepsilon}~.
\end{eqnarray}
The phaseshifts $\delta_l$ are then obtained from the on-shell t-matrix using the
relation
\begin{eqnarray}
  \label{eq:def:phaseshift}
  t_l(p,p)=-\frac{2 p e^{i\delta_l}\sin\delta_l}{m \pi}~,
\end{eqnarray}
where we will concern ourselves only with the $l=0$ phaseshifts.

\begin{figure}[t]
 \begin{center}
  \includegraphics*[width=3.5in]{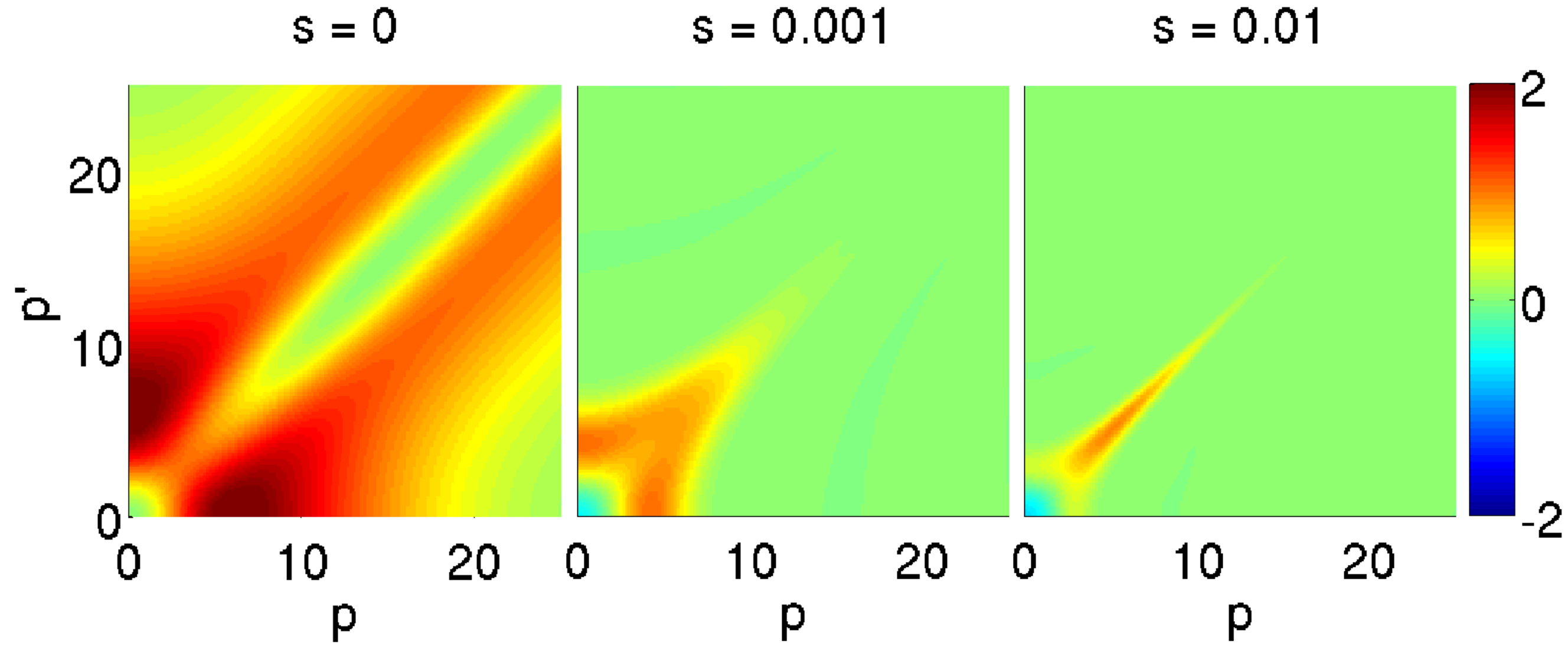}
  \includegraphics*[width=3.5in]{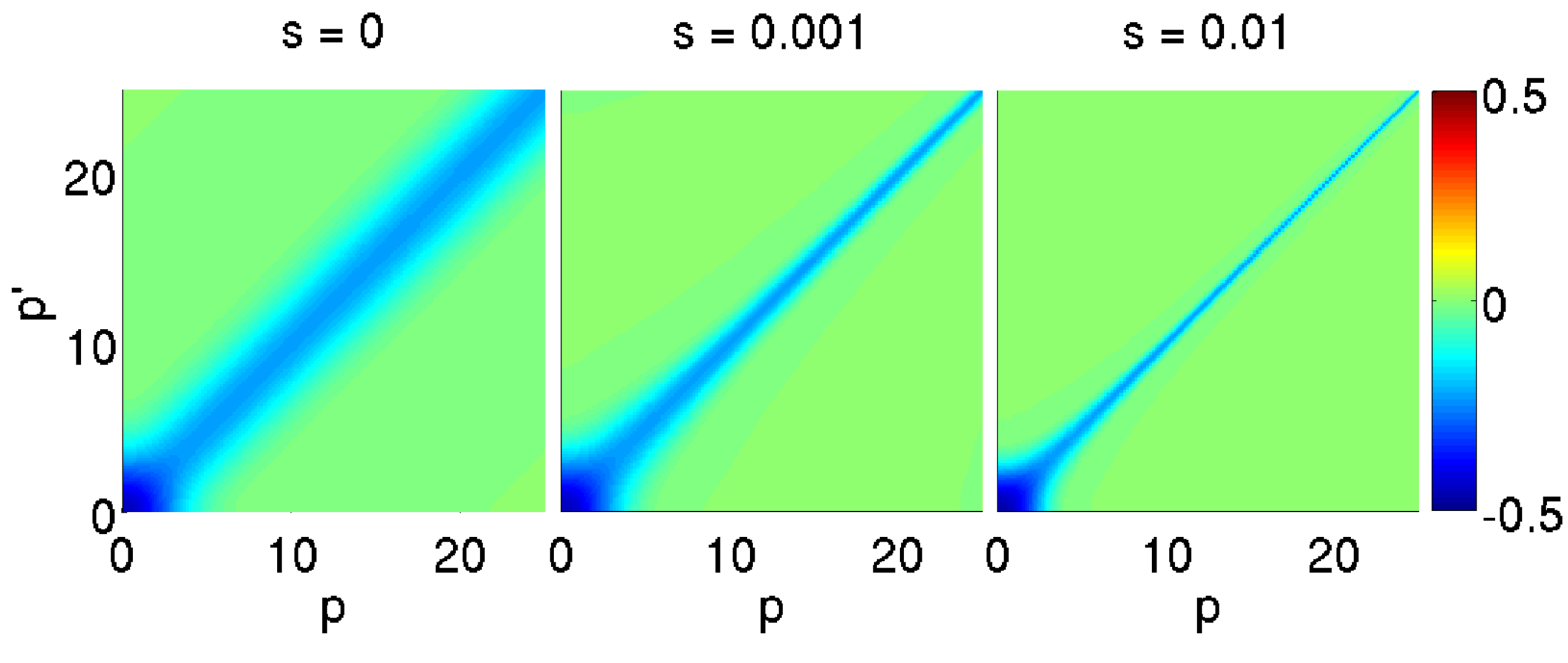}
  \includegraphics*[width=3.5in]{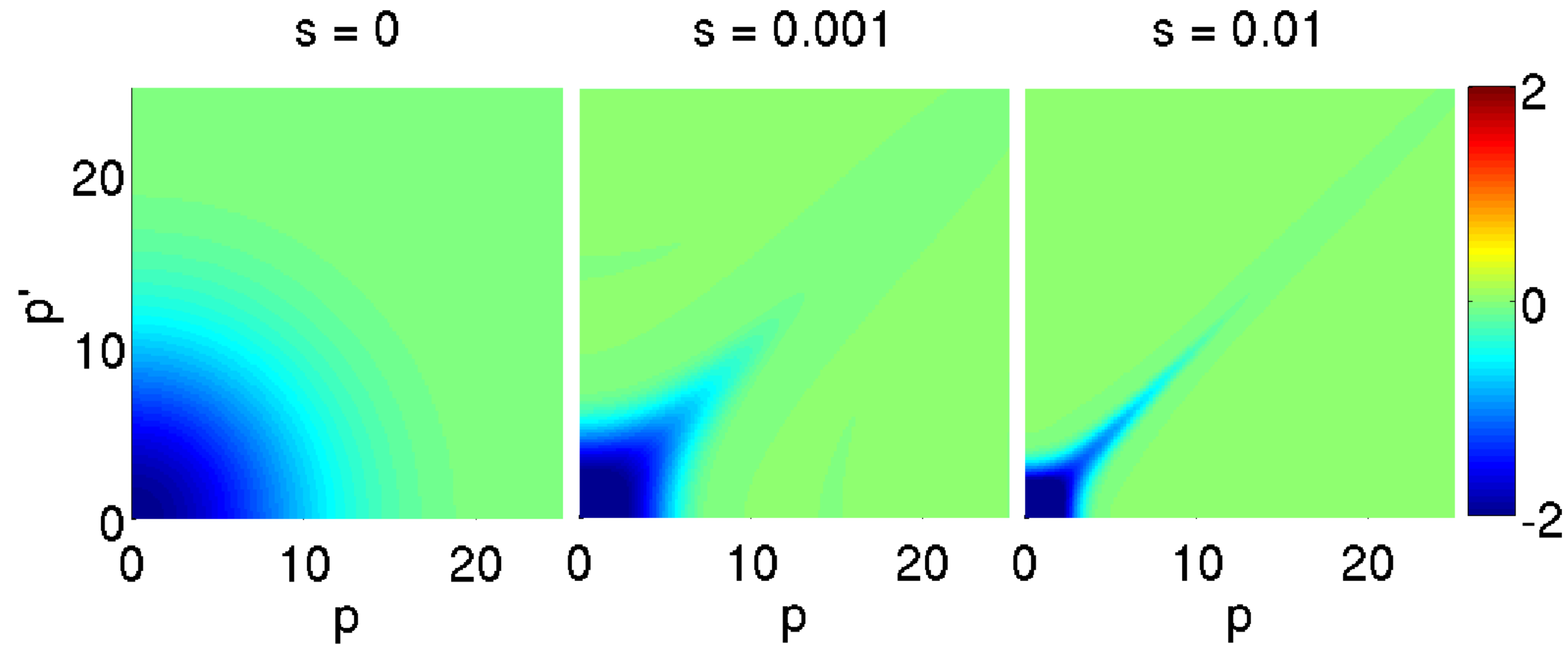}
 \end{center}
 \caption{SRG evolution of the potentials $V_\alpha$ (upper panel),
   $V_\beta$ (middle panel) and $V_{\rm sep}$ (lower panel) as defined
   in Eqs.~\eqref{eq:Valpha} and \eqref{eq:separable} and
   Table~\ref{tab:jurgparam}. The parameters in the separable
   potential are given by $g = -1$ and
    $\Lambda = 10$.
    \label{fig:2bodysrgevolution}}
\end{figure}

\begin{table}[t]
\begin{center}
  \caption{Binding energy $B_2$ for two particles interacting through the
    evolved potential $V_{\rm sep}$ for different $s$. The original potential
    is given by Eq.~\eqref{eq:separable} with $g = -1$ and
    $\Lambda = 10$.}
\label{tab:V1}
\begin{tabular}{c|c}
s &  $B_2$ \\ \hline
0 &  5.12880487\\
0.005 & 5.12880468 \\
0.01 & 5.12880466 \\
0.02 & 5.12880462 \\
0.04 & 5.12880457 \\
0.06 & 5.12880454
\end{tabular}
\end{center}
\end{table}

\begin{figure}[t]
 \begin{center}
  \includegraphics*[width=3.3in]{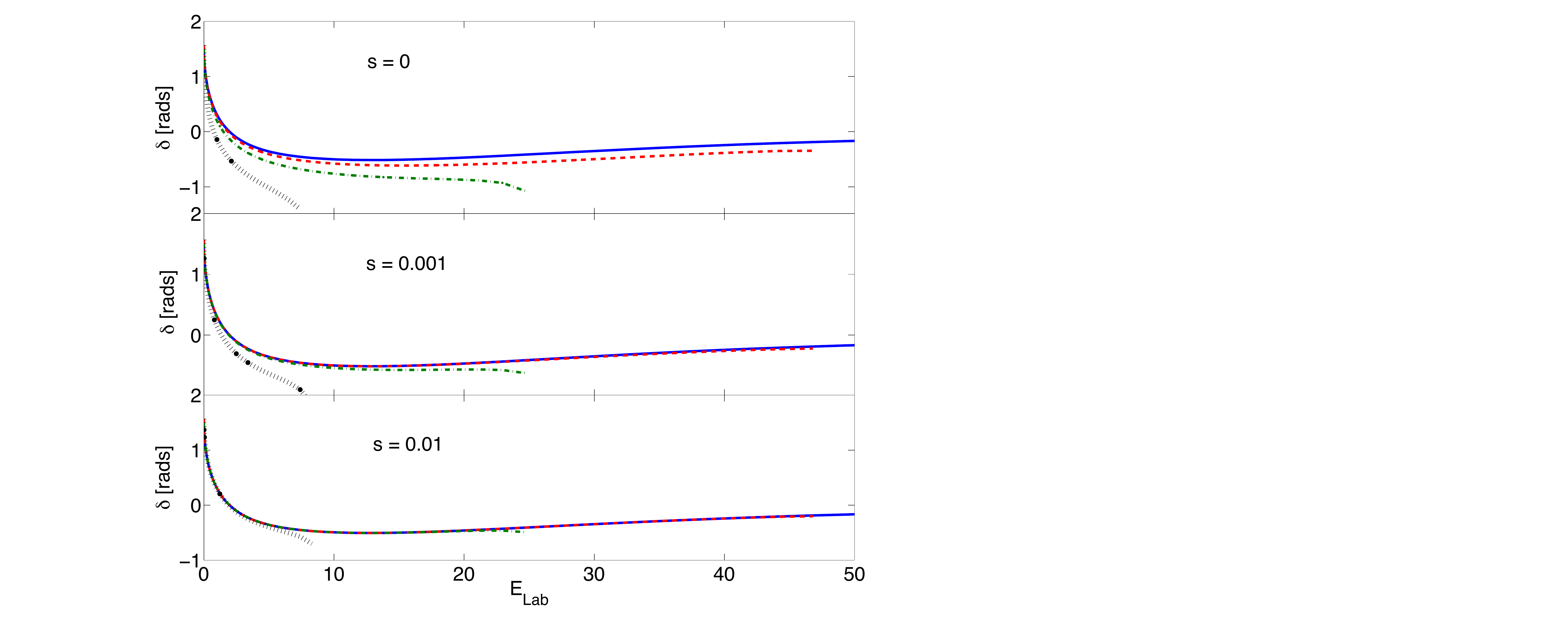}
 \end{center}
 \caption{Two-body scattering phaseshifts calculated from the evolved and unevolved
potential $V_{{\alpha}}$. The different panels correspond to different
values of the evolution parameters $s$ with $s=0$ (upper panel),
$s=0.001$ (middle panel) and $s=0.01$ (lower panel). The different curves denote
results obtained with different values of the Lippmann-Schwinger momentum
space cutoff $\Lambda_{\rm cut}$. The solid line denotes the uncut
result. The dashed, dot-dashed, dotted lines give the result for
$\Lambda_{\rm cut}=7, 5, 3$, respectively.
\label{fig:phase}}
\end{figure}

\begin{figure}[t]
 \begin{center}
  \includegraphics*[width=3.3in]{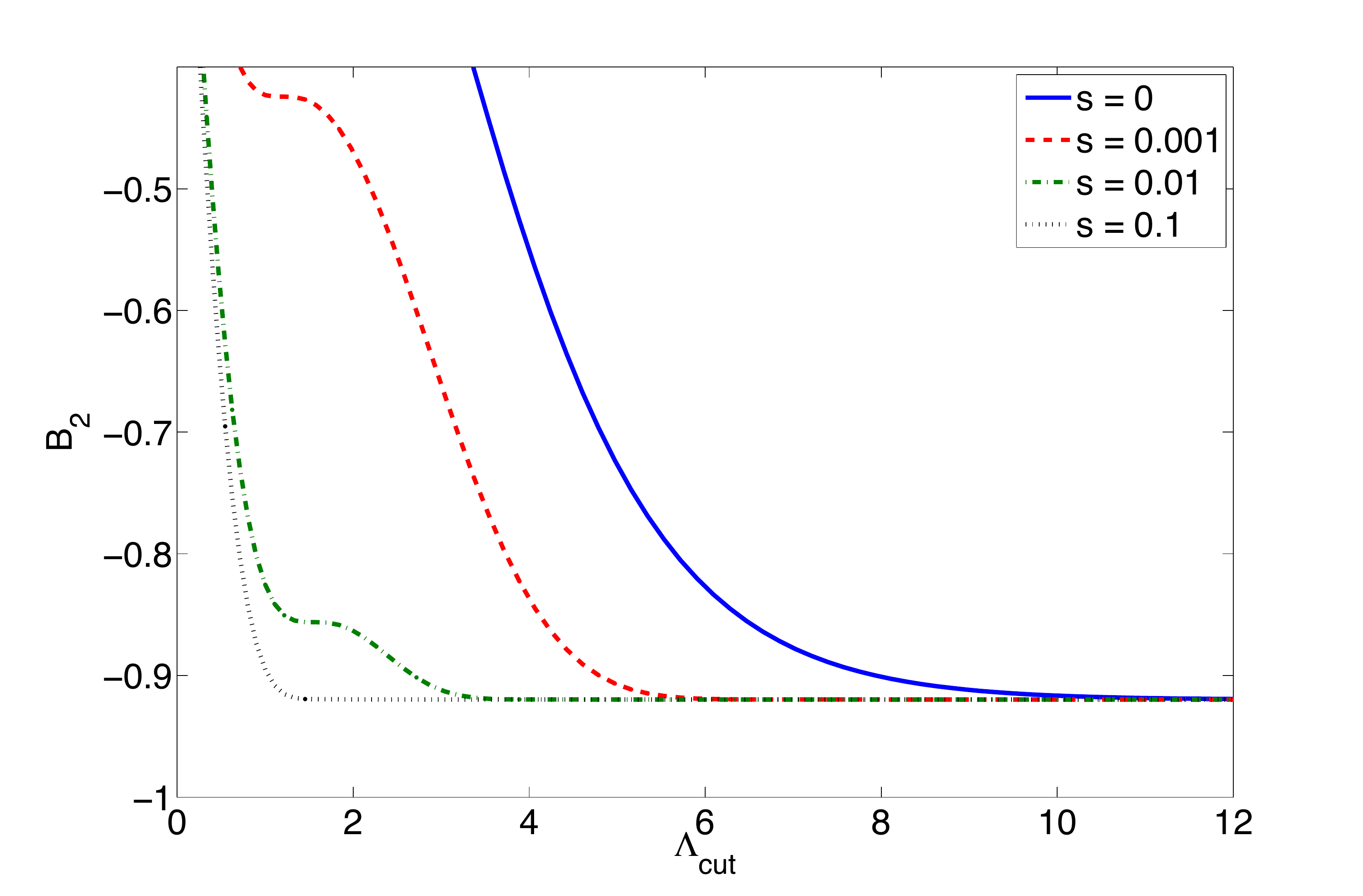}
 \end{center}
 \caption{Two-body binding energy of the evolved potential
   $V_\alpha$ as defined in Eq.~\eqref{eq:Valpha} and
   Table~\ref{tab:jurgparam}. The y-axis denotes the momentum-space
   cutoff $\Lambda_{\rm cut}$ in the momentum-space Schr\"odinger
   equation. The different lines correspond to different values of the
   evolution parameter $s$ with $s=0$ (solid line), $s=0.001$ (dashed
   line), $s=0.01$ (dot-dashed) and $s=0.1$ (dotted line).
   \label{fig:2bodydecoupling}}
\end{figure}
\subsection{Three-Body Observables}
\label{sec:three-body-observ}
In the three-body sector we will focus on binding energies. An
effective way to calculate three-body binding energies in momentum
space is provided by the Faddeev equation.  
\begin{equation}
|\psi\rangle = G_0t_2P|\psi\rangle +
G_0t_2G_0t_3(1+P)|\psi\rangle, \label{eq:fadfinal} 
\end{equation}
where $t_2$ and $t_3$ are the transition operators~\eqref{eq:LSeq}
obtained from the two- and three-body potential terms. 

We can then obtain the total wavefunction by
\begin{equation}
|\Psi\rangle=(1+G_0t_3)(1+P)|\psi\rangle\label{eq:fullwavefunction}.
\end{equation}

\section{Results}
We have implemented the evolution of the two- and three-body
potentials in Python and \textsc{Matlab}. Equations
\eqref{eq:two_body_momspace} and \eqref{eq:srg3body} were discretized
and written as a number of matrix multiplications. They were then
solved using one of the standard ode solvers available in the
corresponding programming language. We did not encounter any stiffness
when solving the differential equations.

We have chosen to work in units where $\hbar^2/m=1$ and all results
are given according to this convention.
The results for two-body binding energies obtained from unevolved
potentials are shown in Table~\ref{tab:jurgtab} and the first row of
Table~\ref{tab:V1}. Three-body binding energies are presented in
Table~\ref{tab:jurgtab}. We note that our results for $V_\beta + V_3$
differ slightly from the values given in Table II of
Ref.~\cite{Jurgenson:2008jp}. However, our results have been
reproduced recently \cite{Kyle:private} and are therefore assumed to
be correct.

\paragraph{Two-Body Evolution/Decoupling:}
We have evolved all previously defined two-body interactions. The
diagonalization as a result of the evolution in the plane that is
spanned by incoming and outgoing relative momenta is common to all
starting interactions. In Fig.~\ref{fig:2bodysrgevolution} we show the
SRG transformation of interactions $V_\alpha$, $V_\beta$ and $V_{\rm
  sep}$, respectively. With increasing flow-parameter $s$ the
potential becomes increasingly more diagonal. The area in the
low-momentum region of these figures that does not get diagonal with
the evolution indicates the presence of a low-momentum two-body bound
state. This is a typical feature of the $T_{\rm rel}$ generator and is
not be present for all other possible generators such as the Wegner
generator used in Ref.~\cite{Wendt:2011qj}.

\begin{figure}[t]
 \begin{center}
  \includegraphics*[width=3.3in]{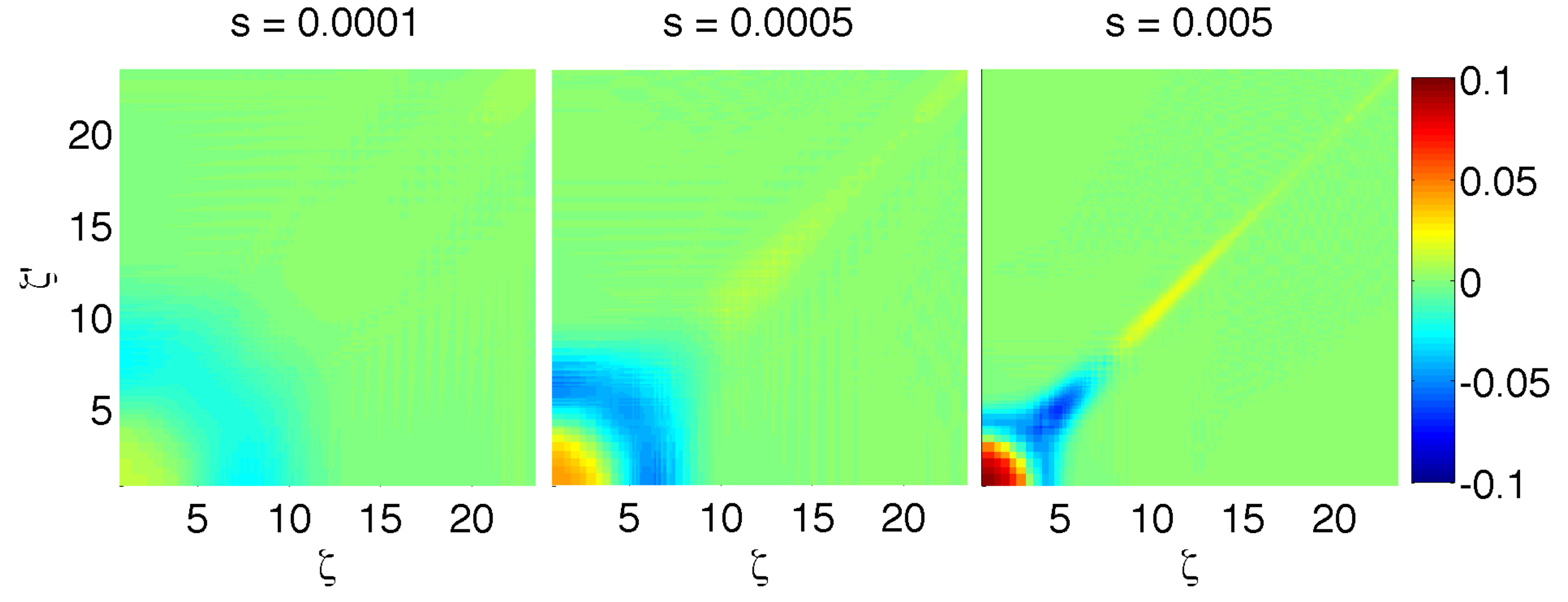}
  \includegraphics*[width=3.3in]{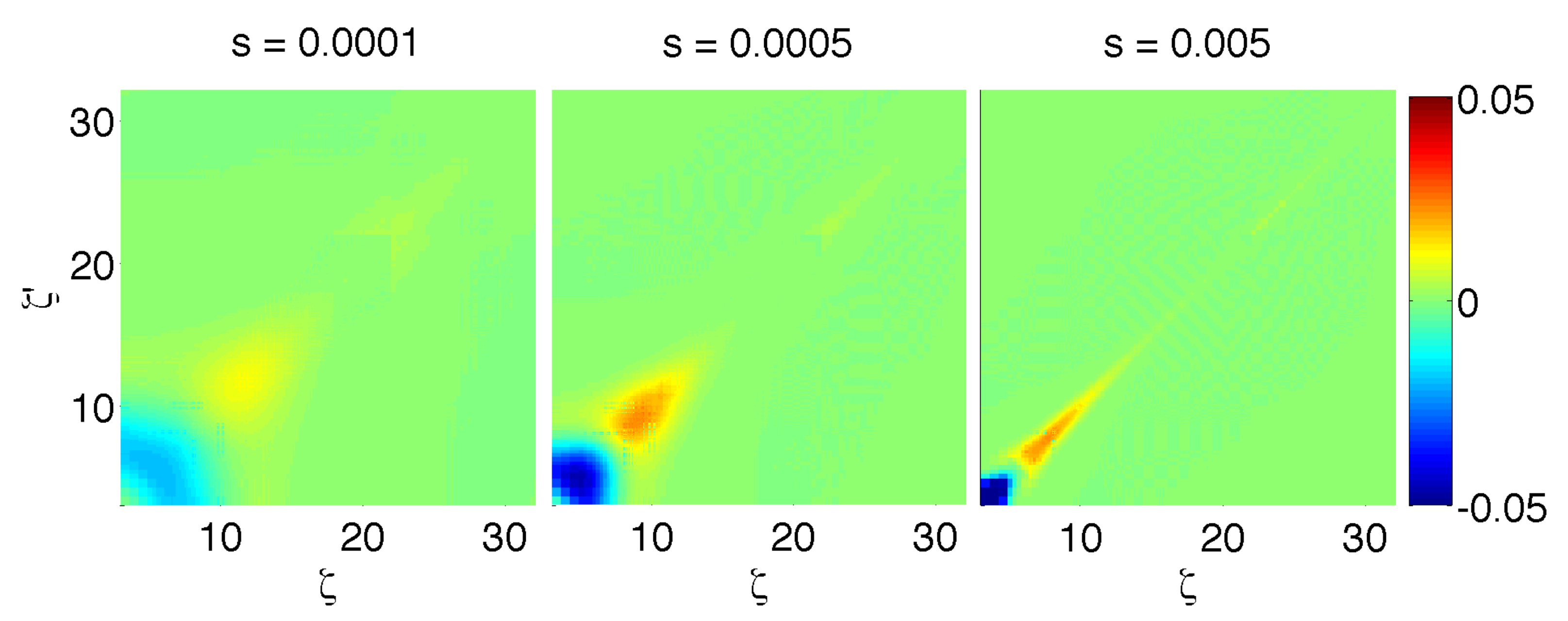}
  \includegraphics*[width=3.3in]{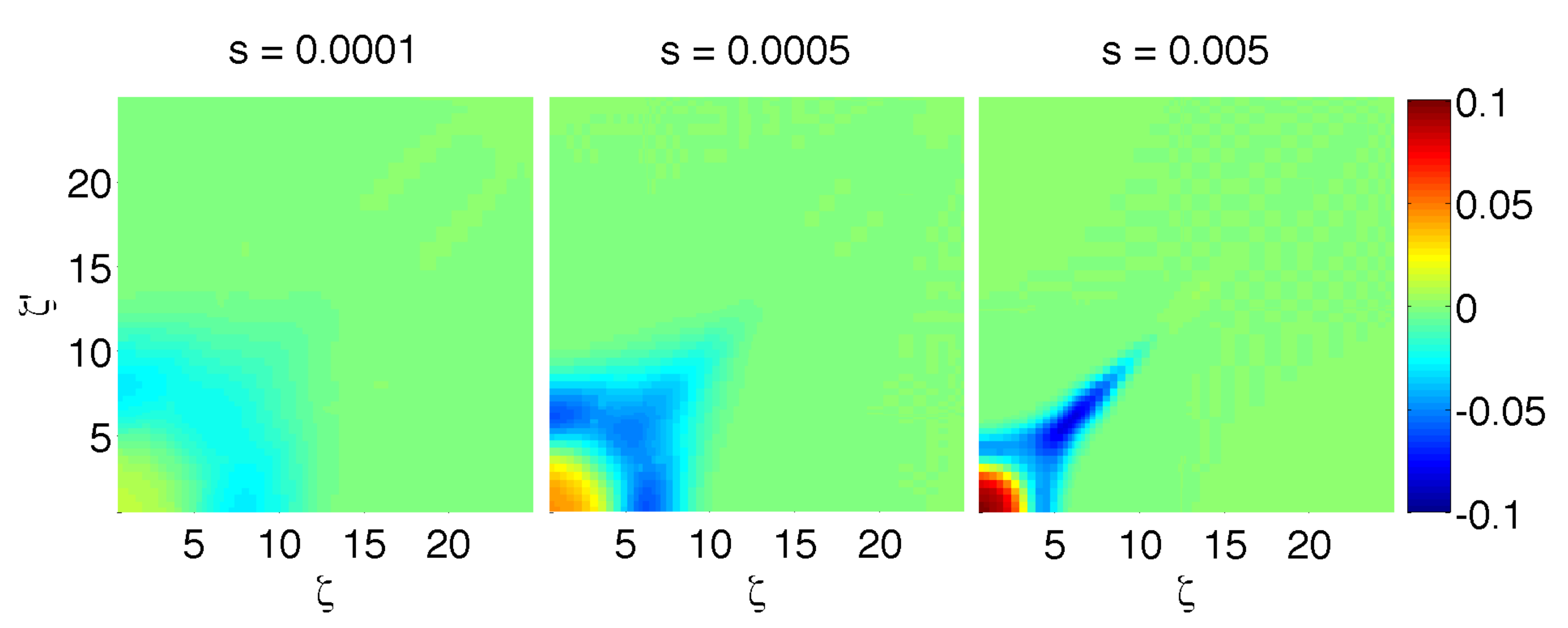}
 \end{center}
 \caption{The induced three-body potential as a function of the inital-
   and final-state hypermomentum $\zeta$ and
   $\zeta'$ for the hyperangles $\theta=\pi/12$ (upper panel), $\theta=\pi/4$
   (middle panel), $\theta=\pi/2$ (lower panel). The
   starting two-body potential is $V_\alpha$, as defined in
   Eq.~\eqref{eq:Valpha} and 
   Table~\ref{tab:jurgparam}. 
   \label{fig:3bodysrgevolution}}
\end{figure}

\begin{figure}[t]
 \begin{center}
  \includegraphics*[width=3.3in]{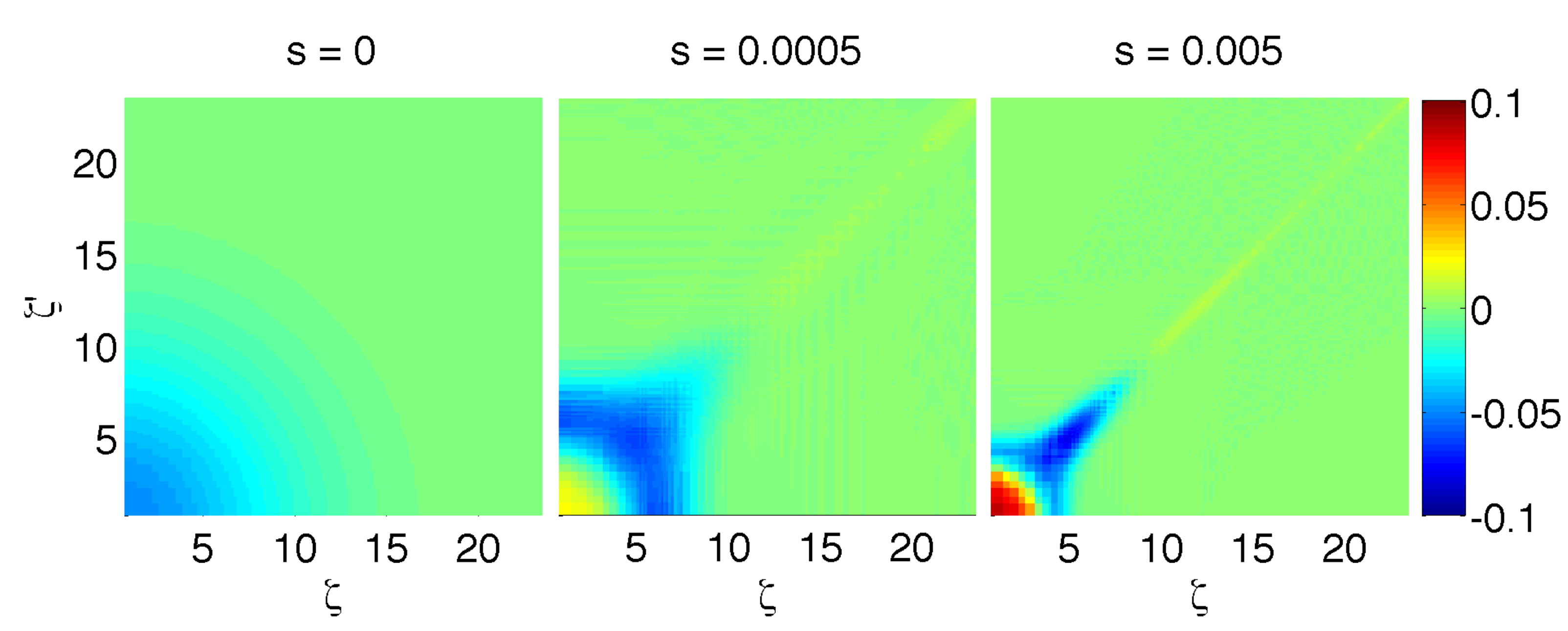}
  \includegraphics*[width=3.3in]{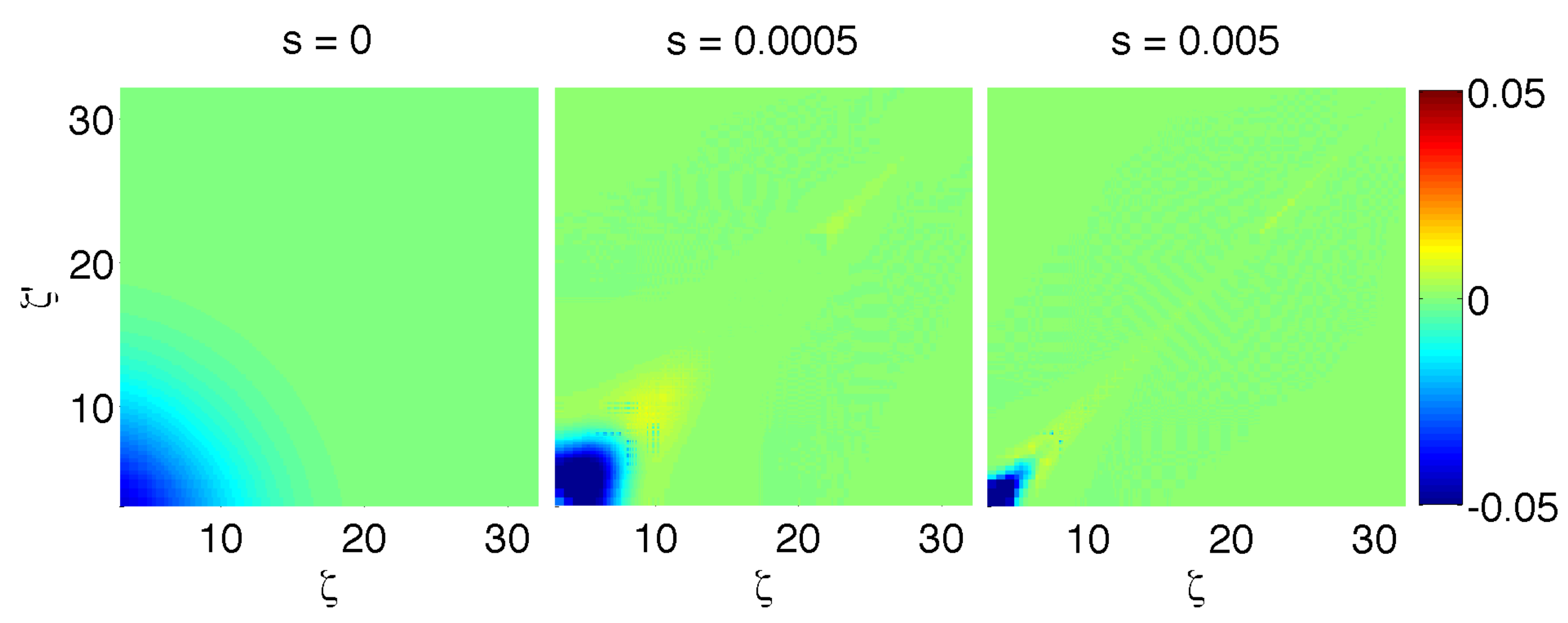}
  \includegraphics*[width=3.3in]{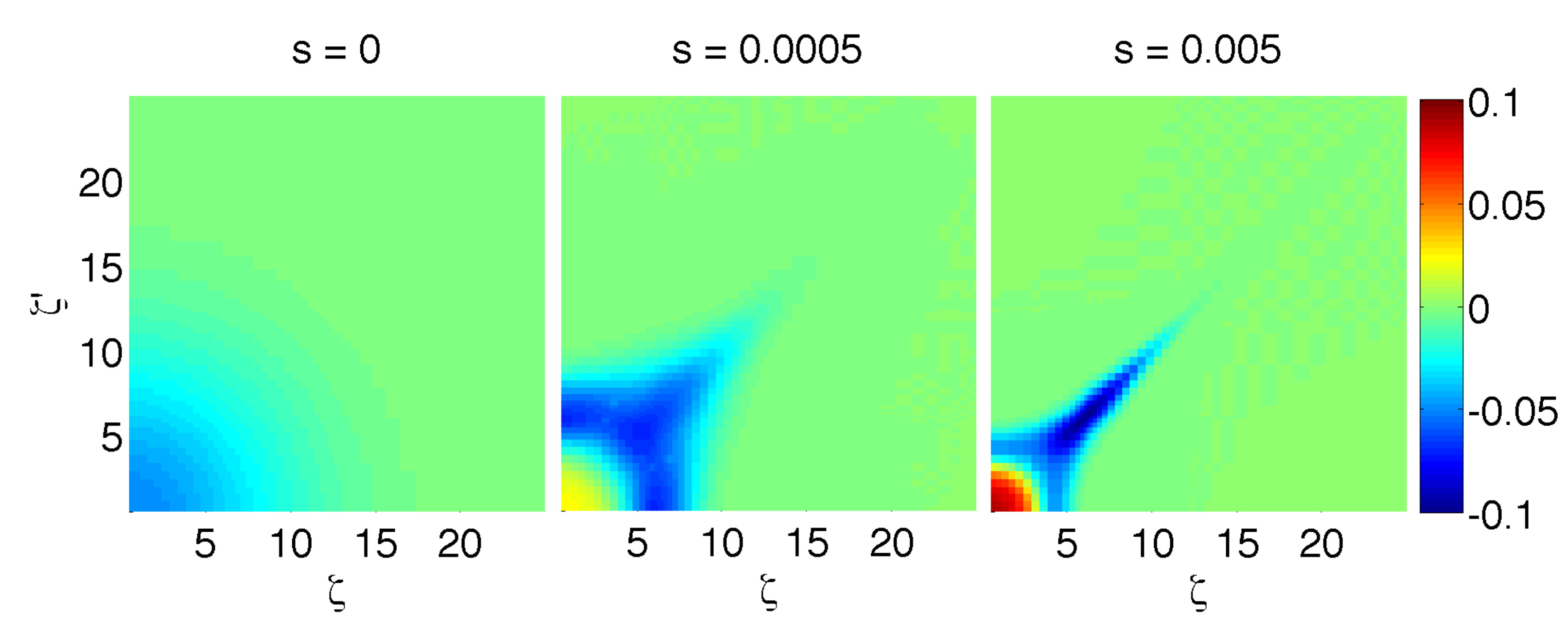}
 \end{center}
 \caption{The evolved three-body potential as a function of the inital-
   and final-state hypermomentum $\zeta$ and
   $\zeta'$ for the hyperangles $\theta=\pi/12$ (upper panel), $\theta=\pi/4$
   (middle panel), $\theta=\pi/2$ (lower panel). The
   starting two-body potential is $V_\alpha$ as defined in Eq.~\eqref{eq:Valpha} and
   Table~\ref{tab:jurgparam}, the starting three-body interaction is
   given in Eq.~\eqref{eq:3body}, where $c_E=-0.05$. 
   \label{fig:3bodysrgevolution-2}}
\end{figure}

The decoupling of large and small momenta in the potential through the
SRG evolution is one of the most important features that result from
employing the diagonal $T_{\rm rel}$ generator. Low-energy observables
can therefore be calculated correctly with a decreased momentum-space
cutoff after sufficient evolution. We illustrate this important
feature of the SRG in Figs.~\ref{fig:phase} and
\ref{fig:2bodydecoupling} where we plot the phaseshifts and two-body
binding energy (obtained with $V_\alpha$) as a function of a sharp
cutoff $\Lambda_{\rm cut}$ in the momentum-space Schr\"odinger
equation for different values of $s$. With increasing flow parameter
$s$, observables become less sensitive to such a truncation indicating
thereby the decoupling of small from large momenta in the potential.
\begin{table}[t]
\begin{center}
  \caption{Two- and three-body binding energies $B_2$ and $B_3$ for
    the starting two-body potentials $V_\alpha$ and $V_\beta$ and
    varying strength of the $s=0$ three-body interaction.}
\label{tab:jurgtab}
\begin{tabular}{cc|cc}
$V_2$&$c_E$&$B_2$&$B_3$\\\hline
$V_\alpha$&-0.10&-0.920&-3.226\\
$V_\alpha$&-0.05&-0.920&-2.885\\
$V_\alpha$&0.00&-0.920&-2.567\\
$V_\alpha$&0.05&-0.920&-2.279\\
$V_\alpha$&0.10&-0.920&-2.027\\
$V_\beta$&-0.10&-0.474&-2.570\\
$V_\beta$&-0.05&-0.474&-2.132\\
$V_\beta$&0.00&-0.474&-1.708\\
$V_\beta$&0.05&-0.474&-1.307\\
$V_\beta$&0.10&-0.474&-0.952\\
\end{tabular}
\end{center}
\end{table}
\paragraph{Three-Body Evolution/Decoupling:}
We have evolved all potentials and obtained not only the corresponding
two-body potential but also the induced three-body interaction
term. Our generator diagonalizes the Hamiltonian in terms of the total
kinetic energy. In Figs.~\ref{fig:3bodysrgevolution} and
\ref{fig:3bodysrgevolution-2}, we have therefore chosen to plot the
three-body interaction in the plane of incoming and outgoing
hypermomentum $\zeta$ and $\zeta'$ for three different hyperangles
$\theta \equiv \arctan \left(\sqrt{3} p / (2 q) \right)$. The induced
three-body potential obtained in the absence of an initial three-body
force is shown in Fig.~\ref{fig:3bodysrgevolution}. Note that, in this
case, the three-body potential is identical to zero for $s=0$. It can
be seen clearly how the evolution induces a three-body potential and
that the strength of off-diagonal elements in hypermomentum space
depends strongly on the flow parameter $s$. The three-body potential
is weak for small $s$ parameter but also couples strongly off-diagonal
elements in the hypermomentum plane. Then, when $s$ increases the
three-body potential becomes stronger but also more diagonal in the
hypermomentum plane. In Fig.~\ref{fig:3bodysrgevolution-2} we show the
same evolution of the three-body interaction where the Hamiltonian
contains the initial three-body force as defined in
Eq.~\eqref{eq:3body} with $c_E=-0.05$.

\begin{figure}[t]
 \begin{center}
  \includegraphics*[width=3.3in]{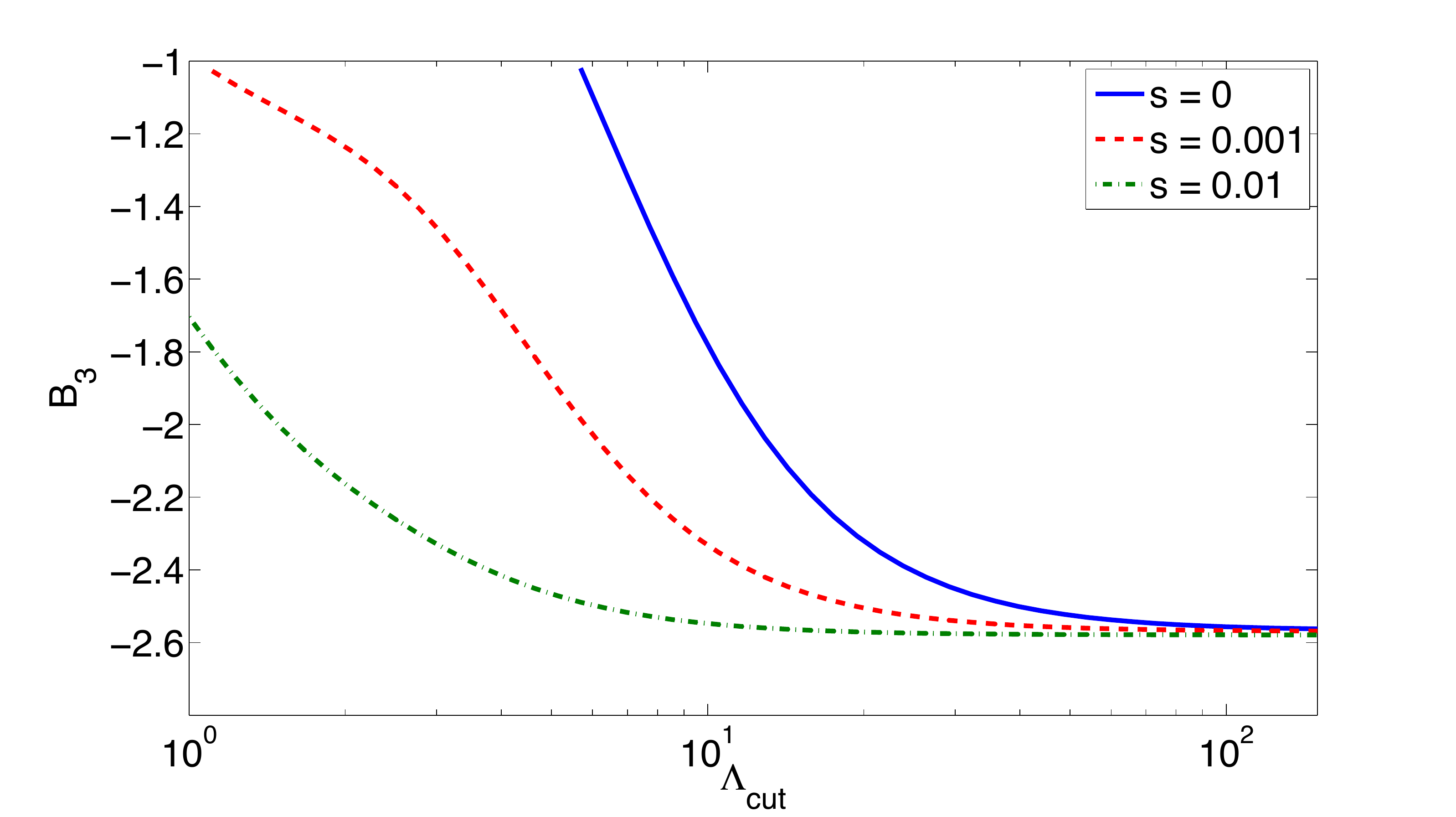}
 \end{center}
 \caption{Three-body binding energy of the evolved potential
   $V_\alpha$ as defined in Eq.~\eqref{eq:Valpha} and
   Table~\ref{tab:jurgparam} and no initial three-body force.
The y-axis denotes the truncation
   cutoff $\Lambda_{\rm cut}$ defined in
   Eq.~\eqref{eq:3bodytruncation}. The different lines correspond to
   different values of the evolution parameter $s$ with $s=0$
   (solid line), $s=0.001$
   (dashed line) and $s=0.01$
   (dot-dashed line).
   \label{fig:3bodydecoupling}}
\end{figure}

Furthermore, we have tested decoupling in the three-body sector
numerically. This was achieved by multiplying the two- and three-body
potentials with regulators that are functions of the two- and
three-body kinetic energy, respectively. In particular, we modify the
evolved potential
\begin{equation}
\label{eq:3bodytruncation}
V_s(p,q,p',q')\rightarrow \exp(-(\zeta^2+\zeta'^2)/\Lambda_{\rm cut}^2) V_s(p,q,p',q')
~.
\end{equation}
In Fig.~\ref{fig:3bodydecoupling} we show the three-body binding
energy as a function of $\Lambda_{\rm cut}$ for different values of the
flow parameter $s$. The longer evolution, measured by $s$,
the smaller is the minimal truncation cutoff $\Lambda_{\rm cut}$ for
which the three-body binding energy remains unchanged. The decoupling
of small and large hypermomenta due to the SRG is therefore clearly
visible.

\section{Summary and Outlook}
In this work we have presented results for the SRG evolution of
three-body interactions for one-dimensional, bosonic model systems
performed in a plane-wave basis. We showed that the evolution in the
plane-wave basis is easily implemented and that observables in the
three-body sector remain unchanged when evaluated with the evolved
potentials. We showed explicitly that the SRG induces a three-body
potential that becomes increasingly diagonal in the hypermomentum
plane when evolved to larger flow parameter $s$. We tested decoupling
explicitly by calculating observables with truncated two- and
three-body interactions. Decoupling works well as illustrated for
binding energies and phaseshifts in the two-body sector and for
binding energies in the three-body sector.

This work presents the first step towards a consistent SRG evolution of
two- and three-nucleon interactions in the plane-wave basis. The
extension to three dimensions and to nuclear systems will require
small modifications of the spatial part of the equations plus the
addition of angular momentum, spin and isospin recoupling to the
problem. The latter is a usual part of standard few-body equations and
we do therefore expect these changes to be straightforward. The
immediate benefit of this implementation is the increased consistency
in a calculation of nuclear matter observables. A plane-wave
formulation will also allow to test more reliably the importance of
four-body forces in a description of nuclear systems. It was claimed
recently that these become of increasing importance (after SRG
evolution) in certain nuclei~\cite{Roth:2011ar}. However, the results
in Ref.~\cite{Roth:2011ar} were obtained with a SRG evolution in the
harmonic-oscillator basis that could in principle be associated with
truncation errors. Furthermore, a formulation in a plane-wave basis
opens up the possibility of identifying relevant scattering
observables in the four-body sector where an emerging four-body
interaction should also be visible.

Another possible avenue for the SRG in three dimensions is the Efimov
effect in three-body systems of identical bosons with large scattering
length. It is an open question how the renormalization group limit
cycle, that was found to occur in this problem \cite{Bedaque:1998km},
manifests itself as the SRG parameter is varied and the induced
three-body interaction changes. Decoupling in the three-body sector
might also provide another path to extract universal properties of the
four-body sector. The SRG might be used to remove three-body bound
states and to calculate the universal properties of highly-excited
four-body states without the need of a scattering calculation as done
for example in Ref.~\cite{Deltuva:2010xd}.

\begin{acknowledgement}
  We thank K. Wendt, D. Lee, E. Jurgenson, R. J. Furnstahl for useful
  discussions. This work was supported by the Chalmers eScience
  Center, the Swedish Research Council (CF,LP), and the European
  Research Council under the FP7 (CF).
\end{acknowledgement}

\end{document}